\documentstyle[epsf,epsfig]{europhys}

\def\etal{{\hbox{{\tenit\ et al.\/}\tenrm :\ }}}

\def\stars{\bigskip\centerline{***}\medskip}

\newif\ifboo \boofalse

\begin{document}

\euro{}{}{}{}
\Date{V4.1}
\shorttitle{S. CIUCHI \etal DIFFERENT SCENARIOS FOR ...}

\title{{\Large Different Scenarios for Critical Glassy Dynamics }}

\author{S. Ciuchi\inst{1}\footnote{e-mail: sergio.ciuchi@aquila.infn.it} and 
        A. Crisanti\inst{2}\footnote{e-mail: andrea.crisanti@roma1.infn.it}}
\institute{
\inst{1} Dipartimento di Fisica, Universit\'{a} de L'Aquila and \\
        Istituto Nazionale Fisica della Materia, Unit\`a dell'Aquila\\
	via Vetoio, 67100 Coppito-L'Aquila, Italy. \\
\inst{2} Dipartimento di Fisica, Universit\'{a} di Roma 
	``La Sapienza'' and \\
        Istituto Nazionale Fisica della Materia, Unit\`a di Roma\\
	P.le A. Moro 2, 00185 Roma, Italy.
}

\rec{}{}

\pacs{
      \Pacs{64.70}{Pf}{Glass transitions}
      \Pacs{75.10}{Nr}{Spin-glass and other random models}
      \Pacs{61.20}{Gy}{Theory and models of liquid structure}
}

\maketitle

\begin{abstract}
We study the role of different terms in the $N$-body potential
of glass forming systems on the critical dynamics near the glass transition.
Using a simplified spin model with quenched disorder, where the
different terms of the real $N$-body potential are mapped 
into multi-spin interactions, we identified three possible scenarios.
For each scenario we introduce a ``minimal'' model representative of the
critical glassy dynamics near, both above and below, the critical transition line.
For each ``minimal'' model we discuss the low temperature equilibrium 
dynamics.
\end{abstract}

In the last years many efforts have been devoted to study the relaxation
dynamics of undercooled liquids 
near the (structural) glass transition.
When the temperature of the liquid is lowered down to the critical 
glass temperature relaxation times becomes exceedingly long and 
diffusional degrees of freedom freeze over very long time scales.
As a consequence
the difference between a structural glass and a disordered system 
with quenched disorder, which may seem essential, becomes less 
and less 
sharp as the transition is approached 
since  the particles in the liquid become trapped in 
random position (cage effect) 
and the dynamics of a single degrees of freedom 
resembles the relaxational dynamics in a random quenched potential.
The idea that a undercooled liquid is a sort of random solid, 
which dates back to Maxwell \cite{Maxwell}, has been recently 
largely used in the study of the glass transition in undercooled liquids.
In this scenario is, for example, the study of the Instantaneous Normal Mode 
(INM) \cite{INM}, where the N-body 
potential is analyzed in terms of normal modes
of oscillations about a given instantaneous configuration.
Using this technique many properties of the N-body potential 
in models of undercooled liquids have been recently
traced out
\cite{INM,Sciort-Kob-Tarta,Peal,Tarta,Keyes-Zurcher}.

In this letter, based on the connection between structural glasses and 
quenched disorder models, we shall analyze from a general point of view the
role of different terms in the $N$-body potential,
such local stress
and other non-harmonic terms, 
on the critical dynamics. 
We are interested only on the general properties therefore
we shall use a simple spin-glass model where the different terms
of the original $N$-body potential are mapped into multi-spin interactions.
In this spirit local stresses are represented by a linear term, 
the harmonic part becomes a two-spin interaction term 
and higher order nonlinear terms becomes $p$-spin interactions with $p>2$.
The use of spin models to study the (structural) glass transition, which 
dates back to the end of 80's \cite{Kirk}, has 
the great advantage that one can construct solvable spin models 
displaying the typical critical behavior of structural glasses
\cite{Kris-pspin-sta,Kris-pspin-dyn,Rev}. 
The model we consider is a spherical spin-glass model with 
(random) potential
\begin{equation}
\label{randomV}
   V[\sigma] = \sum_{p\ge 1} \left( \sum_{1\leq i_1<\cdots < i_p\leq N} 
                V^{(p)}_{i_1 i_2..i_p} 
           \sigma_{i_1}\sigma_{i_2}...\sigma_{i_p} \right)
\end{equation}
where $ V^{(p)}_{i_1 i_2..i_p}$ are uncorrelated zero mean 
Gaussian variables of variance
\begin{equation}
\label{varVp}
   \overline{\left(V^{(p)}_{i_1 i_2..i_p}\right)^2} = 
    \frac{J_p^2 p!}{2N^{p-1}}
\end{equation}
and $\sigma_i$ are $N$ continuous variables obeying
the spherical constraint $\sum_i \sigma_i^2 = N$.
The parameters $J_p$ define the relative strength of the various terms
in (\ref{randomV}) and
can be tuned to reproduce
some of the properties of the $N$-body potential of undercooled liquids 
obtained for instance by INM analysis.
This model has been discussed to some extent in literature, see e.g. Refs. 
\cite{Ref1,Teo,Pari,Rev}, as prototype mean-field model for the structural 
glass transition, and recently also to fit experimental data 
\cite{Krakoviack}. It is known that the high temperature
phase 
is described by the (schematic) Mode Coupling Theory (MCT) 
for structural glasses.
However, to our knowledge, a systematic analysis
of the relation between the leading terms in the potential (\ref{randomV})
and the critical dynamical behavior, similar to what done in MCT
\cite{Gotze-Siogren}, was never done. 
Our analysis identifies three possible scenarios for critical dynamics,
similar to what found in MCT, and allows for the introduction of a 
``minimal'' spin-glass model for each scenario which describes the 
critical dynamics near, both above and below, the transition.
 We stress that while the 
high temperature phase is similar to what found in MCT, the low temperature 
phase is different. 
We finally note that our analysis also reveals all the possible scenarios
for glassy transition that can be obtained with this model.

The relaxation dynamics is defined as usual by the Langevin equation
\begin{equation}
\label{model}
\partial_t\sigma_i = -R\,\sigma_i
                -\frac{\partial\beta V[{\bf \sigma}]}{\partial\sigma_i}
                +\xi _i (t)
\end{equation}
where $\xi_i(t)$ is a Gaussian random field 
(thermal noise) with variance
$\langle \xi_i(t) \xi_i(t')\rangle = 2\,\delta(t-t')$,
 $T=1/\beta$ the temperature and $R$ a Lagrange multiplier to 
ensure the spherical constraint which must be fixed self-consistently.
If only the linear term $p=1$ and one term with $p>2$ are present in 
the expansion (\ref{randomV}) the model is equivalent 
to the spherical $p$-spin 
model introduced in Ref. \cite{Kris-pspin-sta} whose dynamics has been
studied in Ref. \cite{Kris-pspin-dyn,CKu}. 
The case where only the quadratic term $p=2$ is present has been analyzed in
Refs. \cite{Ciuk,fur}.
It is worth noting that there is a nontrivial difference between the 
dynamics 
for $p=2$ and $p>2$. Indeed while in the former case relaxation is
an orientation process  of the state vector $\{\sigma_i\}$
toward a (doubly degenerate) state \cite{Ciuk}, for $p>2$, similar to
what happens in structural glasses,
relaxation occurs in a free energy landscape characterized by many 
(highly degenerate) local minima \cite{Kris-pspin-dyn,KStap}.
In this spirit small non-harmonic terms change qualitatively the
dynamical properties, therefore  in this letter we always 
assume at least cubic nonlinearities.

The analysis of (\ref{model}) simplifies considerably in the 
thermodynamic limit $N\to\infty$, where the dynamics can be described 
by a set of self-consistent equations involving a single spin only
and the averaged correlation and response functions $C(t,t')$ and $G(t,t')$
\cite{SZ,Kirk,Kris-pspin-dyn,BCKM}.



In the high temperature phase at equilibrium $C$ and $G$ are related
by the fluctuation  dissipation theorem (FDT) 
$G(t,t') = G(t-t')=-\theta(t-t')\partial_t C(t-t')$ 
and the the mean-field dynamical equation simplifies further,
for details see e.g. Refs.  \cite{SZ,Kirk,Kris-pspin-dyn,BCKM}.
The resulting mean-field dynamical equation can be written as,
\begin{equation}
\label{dyn-FDT}
[\partial_t+\overline{r}]\, C(t)+ \int_0^t ds\, \Lambda(t-s)\, \partial_s
C(s) = 1-\overline{r}
\end{equation}
where $\Lambda(t)\equiv\Lambda[C(t)]=\sum_{p\ge 2} \mu_p\, C^{p-1}(t)$,
$\overline{r}=R-\sum_{p\ge 2} \mu_p$, $\mu_p = J_p^2 p /2T^2$, 
and $t$ is now a time difference.
For large $t$ we can define the Edward-Anderson 
order parameter as $\lim_{t\to\infty} C(t) =q_0$ which, taking the 
$t\to\infty$ limit of eq. (\ref{dyn-FDT}), obeys the
equation:
\begin{equation}
\label{eq-stato}
\mu_1+\Lambda[q_0]=\frac{q_0}{(1-q_0)^2}
\end{equation}
The parameter $\overline{r}$ has been eliminated using the
the spherical constraint which now reads $C(0)=1$.
This is the ``replica symmetric'' solution for this model.
Stability analysis reveals that this solution is stable iff
\begin{equation}
\label{eq-stab}
\frac{d\Lambda[q_0]}{dq_0} \leq \frac{1}{(1-q_0)^2}
\end{equation}
the equality being satisfied along the transition line.
Since the linear term in (\ref{randomV}) 
acts as an external field,
it can be shown that if $\mu_1=0$ the only stable solution 
is $q_0=0$ while for $\mu_1 > 0$  we have $0< q_0 < 1$ 
\cite{Kris-pspin-sta}. 
The order parameter $q_0$ is the time-persistent part of
the correlation induced by the variance of the 
local stress, therefore $q_0\not= 0$ is not associated to a glassy phase.
We note that in structural glasses 
has been found \cite{Buchenau} 
that local stresses develop 
a zero mean Gaussian distribution approaching the glass transition,
therefore even if their role is irrelevant well inside the liquid phase,
they can be relevant for the dynamical transition.
To study the relaxation near the transition we introduce the
(rescaled) connected correlation function $\phi(t)$ by writing
$C(t)= q_0+(1-q_0)\phi(t)$ where, from eq. (\ref{dyn-FDT}) and 
(\ref{eq-stato}), $\phi(t)$ obeys the equation
\begin{equation}
\label{dyn-Gotze}
\partial_t \phi(t) +\frac{1}{1-q_0}\,\phi(t)+ 
\int_0^t ds\, \Delta\Lambda(t-s) \, \partial_s
\phi(s) = 0
\end{equation}
with $\phi(0)=1$ and 
\begin{equation}
\Delta\Lambda(t) = \Lambda[C(t)] - \Lambda[q_0] 
                \equiv \sum_{k=1}^\infty \lambda_k \phi^k(t) 
\end{equation}
\begin{equation}
\lambda_k = \frac{(1-q_0)^k}{k!} \frac{d^k }{d q_0^k} \Lambda[q_0].
\end{equation}
This equation has the same structure of the schematic MCT equation 
for glasses considered by G\"otze \cite{Gotze}.
The factor $(1-q_0)^{-1}$, absent in the G\"otze equation, 
just changes the form of the long time solution
near the type B transition (see below). 
The G\"otze equation is recovered for $q_0=0$. 

In deriving eq. (\ref{dyn-Gotze}) we made use of FDT 
so that this equation is appropriate  
{\it only} above the dynamical transition.
In the low temperature phase (spin-glass or glass phase) 
the FDT must be modified \cite{Kris-pspin-dyn,CKu} and,
moreover, non-equilibrium \cite{CKu} and
equilibrium \cite{Kris-pspin-dyn} dynamics
are separated by infinite time scales and described using 
different approaches.

\begin{figure}
\epsfbox{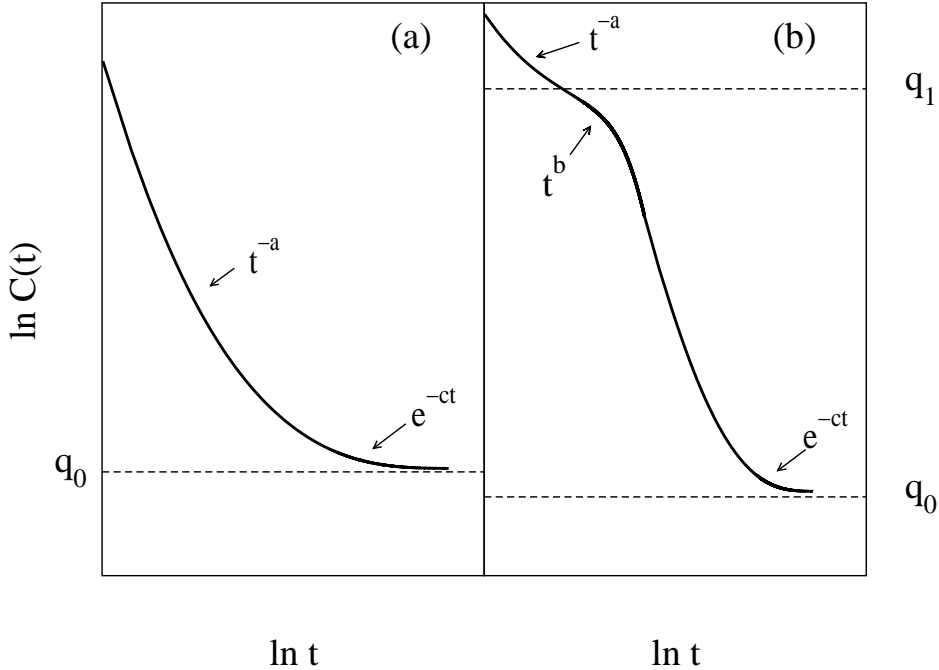}
\caption{Schematic time behavior of the correlation function $C(t)$ near the 
type A transition $(a)$ and the type B transition $(b)$).}
\label{fig-AB}
\end{figure}

The transition from the high to the low temperature phase is of type
A or B depending on the instability of eq. (\ref{dyn-Gotze}). 
The type A transition occurs along the instability line of
solution (\ref{eq-stato}) [equal sign in 
eq. (\ref{eq-stab})] and is the 
analogous of the De Almeida and Thouless line in spin glasses.
Type B transition appears with a sudden appearance of {\it another} 
long time persistent solution: $\lim_{t\to\infty}\phi(t)=f\not=0$ where,
from eq. (\ref{dyn-Gotze}), $f$ is solution of the bifurcation equation:
$\Delta\Lambda[f] = f\,[(1-q_0)(1-f)]^{-1}$.
For $q_0=0$ this reduces to that found by G\"otze \cite{Gotze}.
Defining $f = (q_1 - q_0) (1-q_0)^{-1}$ we recover 
the usual equation for $p$-spin-like models at the
discontinuous transition \cite{Kris-pspin-dyn,BCKM}.
The relaxation dynamics near the two types of transitions is  
quite different, as shown schematically in Fig. \ref{fig-AB}
where relaxation parameters are defined.

Which kind of transition -- type A or B -- takes place depends on
the parameters $\mu_p$.
Near the transition only the first two non-zero terms in the sum 
(\ref{randomV}) are relevant, all others giving sub-leading corrections
which can be neglected at the transition. 
The six non-trivial cases are reported in table \ref{fig-table}. 
All other cases can be qualitatively mapped to one of these.
With $\mu_{p>3}$ we mean a generic non-harmonic term of order $p>3$.
\begin{table}
\begin{tabular}{|c|c|c|c||c|c|c|}
\hline
$\mu_1$ & $\mu_2$ & $\mu_3$ & $\mu_{p>3}$ & $\lambda_1$ & $\lambda_2$ & 
$\lambda_3$ \\
\hline\hline
 0  & 1 & 1 & - & 1 & 1 & - \\ \hline
 1  & 1 & 1 & - & 1 & 1 & - \\ \hline
 1  & 1 & 0 & 1 & 1 & 1 & - \\ \hline 
 1  & 0 & 1 & - & 1 & 1 & - \\ \hline 
 0  & 1 & 0 & 1 & 1 & 0 & 1 \\ \hline
 0  & 0 & 1 & 1 & 0 & 1 & 1 \\ \hline
\end{tabular}
\caption{ 0 = missing, 1 = present , - = irrelevant}
\label{fig-table} 
\end{table}
The six cases can be grouped into three different 
classes depending on the two relevant $\lambda_k$. 
Following a notation introduced by G\"otze, we call them $1-2$, 
$1-3$ and $2-3$, respectively. 
For each class we can define a ``minimal'' spin-glass model obtained 
by retaing only two terms in the sum (\ref{randomV}).
If we denote these modes with ``$x+y$''-SG model, where $p=x,y$
are the two retained terms, 
the simplest choice for spin-glasses ``minimal'' model is: 
$2+3$-SG model for class $1-2$,
$2+4$-SG model for class $1-3$ and
$3+4$-SG model for class $2-3$ [table \ref{fig-table}].
With this choice the relation between $\mu$'s and $\lambda$'s is
very simple: $\lambda_k = \mu_{k+1}$. One could use other choices,
e.g. the $1+3$-SG model for the class $1-2$. This would lead to a different
relation between $\mu$'s and $\lambda$'s,
but will not change the results if expressed in terms of 
$\lambda$'s.

The ``minimal'' models can be analyzed in full details
also far from the transition and hence can be used to gather more 
informations on the low temperature phase in the different classes. 
For this reason in what follow we shall restrict to the ``minimal'' 
models and consider the equilibrium dynamics in the low temperature phase. 
In is clear that the results we shall obtain are also valid
for the full model as far as the extra terms in potential 
(\ref{randomV}) can be neglected, i.e., 
{\it only} near the transition.
The equilibrium dynamics in the low temperature phase
can be studied using the technique of Ref.\cite{SZ,Kris-pspin-dyn}
which assume equilibration  on very long times and a
modified form of FDT for the very slow processes. 
This method is known to reproduce the low temperature phase statics.
The calculation 
is lengthly and will not be reported. Here we quote the main results.
The non-equilibrium aging phenomena can be studied using the 
technique of Ref. \cite{CKu}, but these are beyond the purpose of this 
paper.

\begin{figure}
\epsfbox{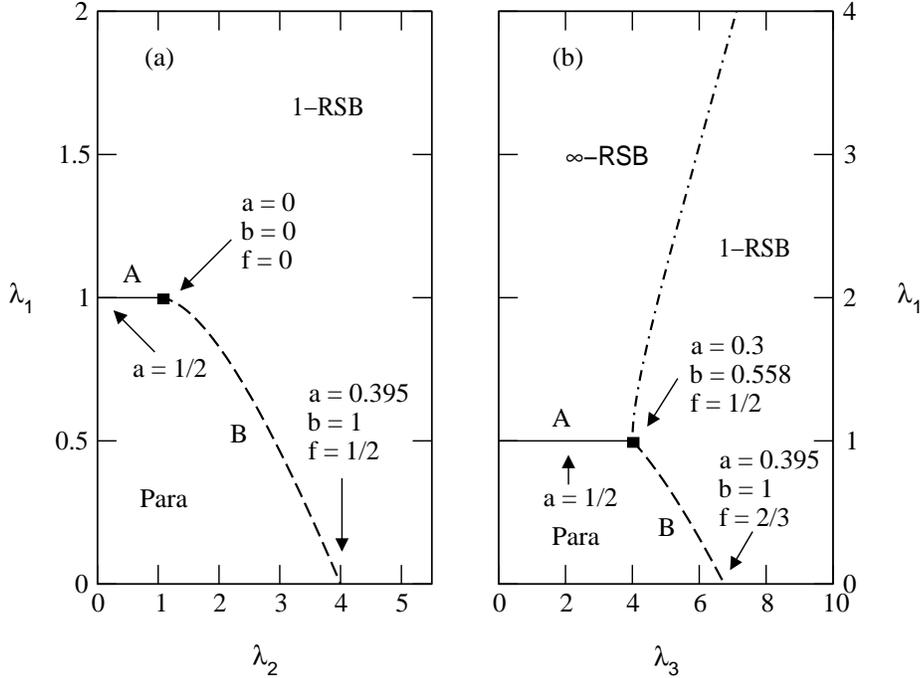}
\caption{The phase diagram of the ``minimal'' $2+3$-SG model $(a)$, and
         of the ``minimal'' $2+4$-SG model $(b)$. 
         The exponents of $C(t)$ [see Fig. \protect\ref{fig-AB}] 
         and the value of $f$
         are shown at the edges of the transition lines.}
\label{fig-2-3}
\end{figure}

In Fig. \ref{fig-2-3} (a) we report the 
phase diagram in the $\lambda_1-\lambda_2$ plane
of the ``minimal'' $2+3$-SG model. This is equal to the phase
diagram of the spherical $p$-spin model in a field 
\cite{Kris-pspin-sta,Kris-pspin-dyn} when expressed in 
terms of $\lambda_{1,2}$.
For small $\lambda_2$ a transition of type A separates a paramagnetic
from a glassy phase described by ``1-step'' replica symmetry 
broken solution.  By increasing $\lambda_2$ along the critical line
$\lambda_1=1$ we eventually reach a tricritical point 
where the type B solution appears and the transition changes from A to B. 
Since $f$ is zero at this point there is no jump in the 
bifurcation parameter. As far as the transition lines are concerned,
this phase diagram is similar to what obtained in MCT for 
structural glasses \cite{Gotze}, however the properties of the low 
temperature phase are different \cite{Kris-pspin-sta,Kris-pspin-dyn}.

The phase diagram of the ``minimal'' $2+4$-SG model, 
shown in Fig. \ref{fig-2-3} (b),  is more reach.
The high temperature phase is still separated from the low temperature
phase by a type A transition for small $\lambda_3$ and by a type B
transition for higher $\lambda_3$. However the low temperature
phase consists now of two different glassy phases: one described by 
``1-step'' and one by ``$\infty$-steps'' replica symmetry broken solution. 
The transition from $1$-step to $\infty$-steps RSB appears as an instability 
of the $1$-step RSB phase -- type A -- and in this respect is similar
to what found for the Ising $p$-spin model \cite{Gard}. 
The point where the type A and
type B transitions meet is not a critical point since the two lines just 
cross, and $f$ jumps discontinuously. 
The possibility of two different glassy phases in model (\ref{randomV})
was discussed
in Ref.\cite{Teo} from considerations on the statics of the model,
but the transition line between the phases was not 
determined. We stress that the observed low temperature phase is 
completely different from what predicted by G\"otze and Sj\"ogren
\cite{Gotze-Siogren}
using the MCT equation (\ref{dyn-Gotze}).
Indeed they find two ``$1$-step'' RSB phases
separated by a type B transition line ending at a critical point.
This, however, does not contradicts our results since as previously
discussed 
in the low temperature phase eq. (\ref{dyn-Gotze})
is not appropriate for our model.

\begin{figure}
\epsfbox{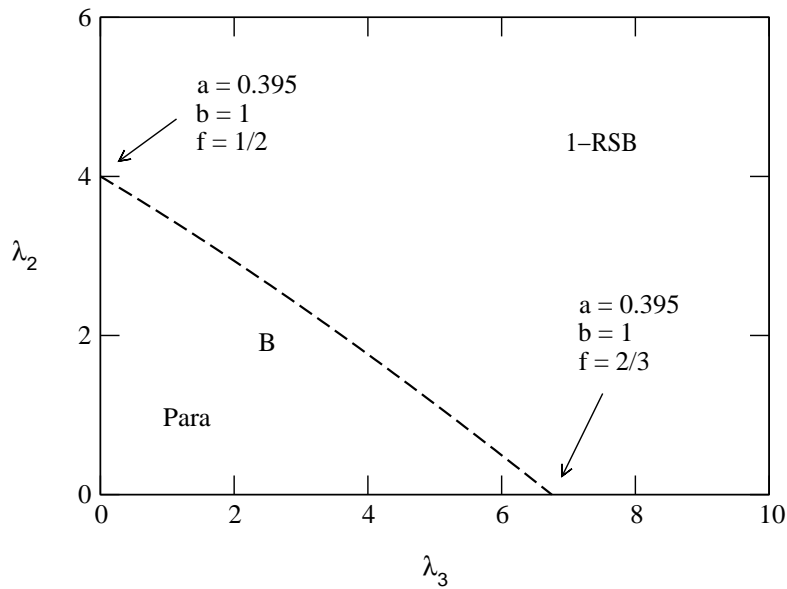}
\caption{The phase diagram of the ``minimal'' $3+4$-SG model.
         The exponents of $C(t)$ [see Fig. \protect\ref{fig-AB}] 
         and the value of $f$
         are shown at the edges of the transition lines.}
\label{fig-3-4}
\end{figure}
The ``minimal'' $3+4$-SG model is the simpler one since the absence 
of quadratic term in  the potential makes the continuous transition 
impossible \cite{Teo}. A discontinuous transition of type B separates 
the high temperature phase form a low temperature phase of 
``1-step'' RSB type, see Fig. \ref{fig-3-4}.

In conclusion in this Letter we have 
shown that simple multi-spin interactions spin-glass spherical models,
which obeys MCT equations above the dynamical transitions, lead
to three different scenarios for critical dynamics according to
the leading terms in the potentials. The type of dynamical transitions can be
continuous as well as discontinuous. From a general point of view a 
continuous transition
may occur only if harmonic and/or linear terms are important.
Within the assumption that different terms of the real $N$-body potential 
of glass-forming systems 
can be mapped into the multi-spin interaction terms in (\ref{randomV}), 
using the informations from 
INM studies the type of scenario for the critical dynamics can be predicted. 

Schematic models of MCT for structural fragile glasses
and undercooled liquids belong to the
$1-2$ class of universality. This is in agreement with our conclusions 
since in these systems local stresses and cubic nonlinearities
are known to be important \cite{Tarta}.
Based on the same assumptions the continuous transition observed 
in the rotational dynamics of linear molecules 
\cite{franosch-dumb,schilling} could be related to the relevance of linear
and/or harmonic terms in the rotational degrees of freedom potential.
To test our conjectures it would be 
also of interest to find physical systems
belonging to $1-3$ or $2-3$ dynamical universality class.

To get more insight on the low temperature phase for each scenario
we have introduced a ``minimal'' SG model and discussed the equilibrium
low temperature dynamics. In particular for the $1-3$ class has been
calculated the critical transition line between the two glassy phases.
The results found for the ``minimal'' models do apply to the full
model near the transition lines, both above and below the transition.
Recent results \cite{Krisanti} have shown that {\it finite-size} mean-field
$p$-spin-like models, where activated processes are allowed, exhibit 
strong similarities with structural fragile glasses. Therefore the 
``minimal'' models can be highly valuable to study
the glass transition in the different scenarios
also beyond MCT. Work in this direction is in progress.

We finally stress that there are examples of glass forming systems where 
the MCT scenario is different form those proposed in the present Letter.
For example a mixture of sticky hard-sphere exhibits a two different
glass phase separated by a type B transition \cite{papertar}.
In this case the short wave-length dependence of vertices in the MCT 
plays a crucial role and therefore a simple model
where inhomogeneous spatial fluctuations are neglected, 
as the one considered in this Letter, cannot be used.

\stars{We wish to thank A. Cavagna, C. Donati, F. Sciortino and P. Tartaglia
for valuable discussions and critical reading of the manuscript.}


\end{document}